\documentclass[runningheads,fleqn]{svmult}

\usepackage{makeidx}   
\usepackage{graphicx}  
\usepackage{subeqnar}  
\usepackage{multicol}  
\usepackage{physproc}  
\makeindex             

\begin{document}

\title*{Non-equilibrium Surface Growth and Scalability of Parallel 
Algorithms for Large Asynchronous Systems}


\titlerunning{Parallel Scalability for Large Asynchronous Systems}

\author{
G. Korniss\inst{1}
\and M.~A. Novotny\inst{1}
\and Z. Toroczkai\inst{2}
\and P.~A. Rikvold\inst{1,3}
}
\authorrunning{Korniss et al.}
%
%
\institute{
School of Computational Science and Information Technology, \\
Florida State University, Tallahassee, Florida 32306-4120, USA
\and
Department of Physics, University of Maryland, \\
College Park, MD 20742-4111, USA
\and
Department of Physics and Center for Materials Research and Technology, \\
Florida State University, 
Tallahassee, Florida 32306-4350, USA
}

\maketitle

\begin{abstract}
The scalability of massively parallel algorithms is a fundamental question
in computer science. We study the scalability and the efficiency of a 
conservative massively parallel algorithm for discrete-event simulations
where the discrete events are Poisson arrivals. The parallel algorithm is 
applicable to a wide range of problems, including dynamic Monte Carlo 
simulations for large asynchronous systems with short-range interactions. 
The evolution of the 
simulated time horizon is analogous to a growing and fluctuating surface,
and the efficiency of the algorithm corresponds to the density of local minima
of this surface. In one dimension we find that the steady state of the 
macroscopic landscape is governed by the Edwards-Wilkinson Hamiltonian, which 
implies that the algorithm is scalable. Preliminary results for 
higher-dimensional logical topologies are discussed.
\end{abstract}

\section{Introduction}

Dynamic Monte Carlo (MC) simulations are invaluable tools
for investigating the evolution of complex systems. 
For a wide range of systems it is plausible to assume (and in rare cases it 
is possible to derive) that attempts to 
update the state of the system form a Poisson process. The basic notion
is that time is continuous, and the {\em discrete events} (update attempts) 
occur instantaneously. The state of the system remains constant between events.
It is worthwhile to note that the standard random-sequential update schemes
(easily implementable on serial computers) produce this dynamics for 
``free:'' the waiting-time distribution for the attempts to update 
each subsystem or component is geometrical and approaches the 
exponential distribution in the large-system limit. This uniquely 
characterizes the Poisson process.

The parallel implementation of these dynamic MC algorithms belongs to the 
class of parallel discrete-event simulation, which is one of the most 
challenging areas in parallel computing \cite{Fuji}. The numerous 
applications range from the natural sciences and engineering to computer 
science and queueing networks. For example, in lattice Ising models the 
discrete events are spin-flip attempts, while in queueing systems they are 
job arrivals. The difficulty of parallel discrete-event simulations is that
update attempts are not synchronized by a global clock. 
In fact, the traditional dynamic MC algorithms were long believed to be 
{\em inherently} serial, i.e., in spin language, the corresponding algorithm 
was thought to be able to update only one spin at a time. However, Lubachevsky 
presented an approach for parallel simulation of these systems \cite{Luba} 
{\em without} changing the underlying Poisson process. 
Applications include modeling of cellular communication
networks \cite{GLNW}, particle deposition \cite{LPR}, and 
metastability and hysteresis in kinetic Ising models \cite{KNR}.

In a distributed massively parallel scheme each processing element (PE)
carries a subsystem of the full system. The parallel algorithm must
{\em concurrently} advance the Poisson streams corresponding to each 
subsystem {\em without} violating causality. This requires the concept of 
{\em local simulated time}, as well as a synchronization scheme. 
Intuitively it is clear that systems with 
short-range interactions contain a ``substantial'' amount of parallelism.
For the ``conservative'' approach \cite{Luba}, the efficiency of the algorithm
is simply the fraction of PEs that are guaranteed to attempt the update 
without breaking causality. The rest of the PEs must idle.

\section{Time Horizon Evolution and Efficiency Modeling}

We consider a $d$-dimensional hypercubic 
regular lattice topology where the underlying 
physical system has only nearest-neighbor (nn) interactions (e.g., Glauber 
spin-flip dynamics) and periodic boundary conditions.  
The scalability analysis is made for the ``worst-case''
scenario in which each PE hosts a single site (e.g., one spin) of the 
underlying system.
While this may be the only scenario for a special-purpose computer with
extremely limited local memory, on architectures with relatively large memory
one PE can host a block of sites. This substantially increases the efficiency,
bringing it to the level of practical applicability \cite{KNR}.

In the basic parallel scheme \cite{Luba}, each PE generates its own 
{\em local simulated time} for the next update attempt. The set of 
local times $\{\tau_{i}(t)\}_{i=1}^{L^d}$ constitute the simulated time 
horizon. Here, $L$ is the linear size of the lattice ($L^d$ is the number of 
PEs), and $t$ is the index of the simultaneously performed parallel steps.
Initially, $\tau_i(0)$$=$$0$ for every site. 
At each parallel time step, {\em only} those PEs for which the local simulated 
time is {\em not greater} than the local simulated times of their nn
can attempt the update and increment their local time by an exponentially 
distributed random amount, $\eta_{i}(t)$. Without loss of generality we
take $\langle\eta_{i}(t)\rangle$$=$$1$. The other PEs idle.
Due to the continuous nature of the random simulated times, for $t$$>$$0$ 
the probability of equal-time updates for any two sites is of measure zero.
The comparison of the nn simulated times and idling if 
necessary enforces causality. Since at worst the PE with the absolute
minimum simulated time makes progress, the algorithm is free from deadlock.
For this basic conservative scheme, the theoretical efficiency (ignoring 
communication overheads) is simply the fraction of non-idling PEs. This 
corresponds to the {\em density of local minima} of the simulated time horizon.
Note that the evolution of the simulated time horizon is {\em completely 
independent} of the underlying model (except for its topology) 
\mbox{and can be written as:}
\begin{equation}
\tau_{i}(t+1) =   \tau_{i}(t)
 + \prod_{j\in D^{\rm nn}_{i}} \Theta\left(\tau_{j}(t)-\tau_{i}(t)\right) 
\eta_{i}(t) \;. \label{tau_evolution}
\end{equation}
Here $D^{\rm nn}_{i}$ is the set of nearest neighbors of 
site $i$, and $\Theta(\cdot)$ is the Heaviside step function.
The evolution of the simulated time horizon is clearly analogous to an
irreversibly growing and fluctuating surface.

There are two important quantities to study. The first is 
the density of local minima, $\langle u(t)\rangle_L$, in particular its
asymptotic (or steady-state) value and finite-size effects. 
It corresponds directly to the efficiency of the algorithm.
The second is the surface width, $\langle w^2(t)\rangle$$=$$(1/L^d) 
\langle
\sum_{i=1}^{L^d}\left[\tau_i(t)-\overline\tau(t)\right]^2 
\rangle$, where $\overline\tau(t)$$=$$(1/L^d)\sum_{i=1}^{L^d}\tau_i(t)$.
It describes the macroscopic roughness of the time horizon and
has important consequences for actual implementations \cite{GSS} 
(e.g., optimal buffer size for a collecting statistics network \cite{Luba}).

\begin{figure}[t]
\includegraphics[width=.3\textwidth]{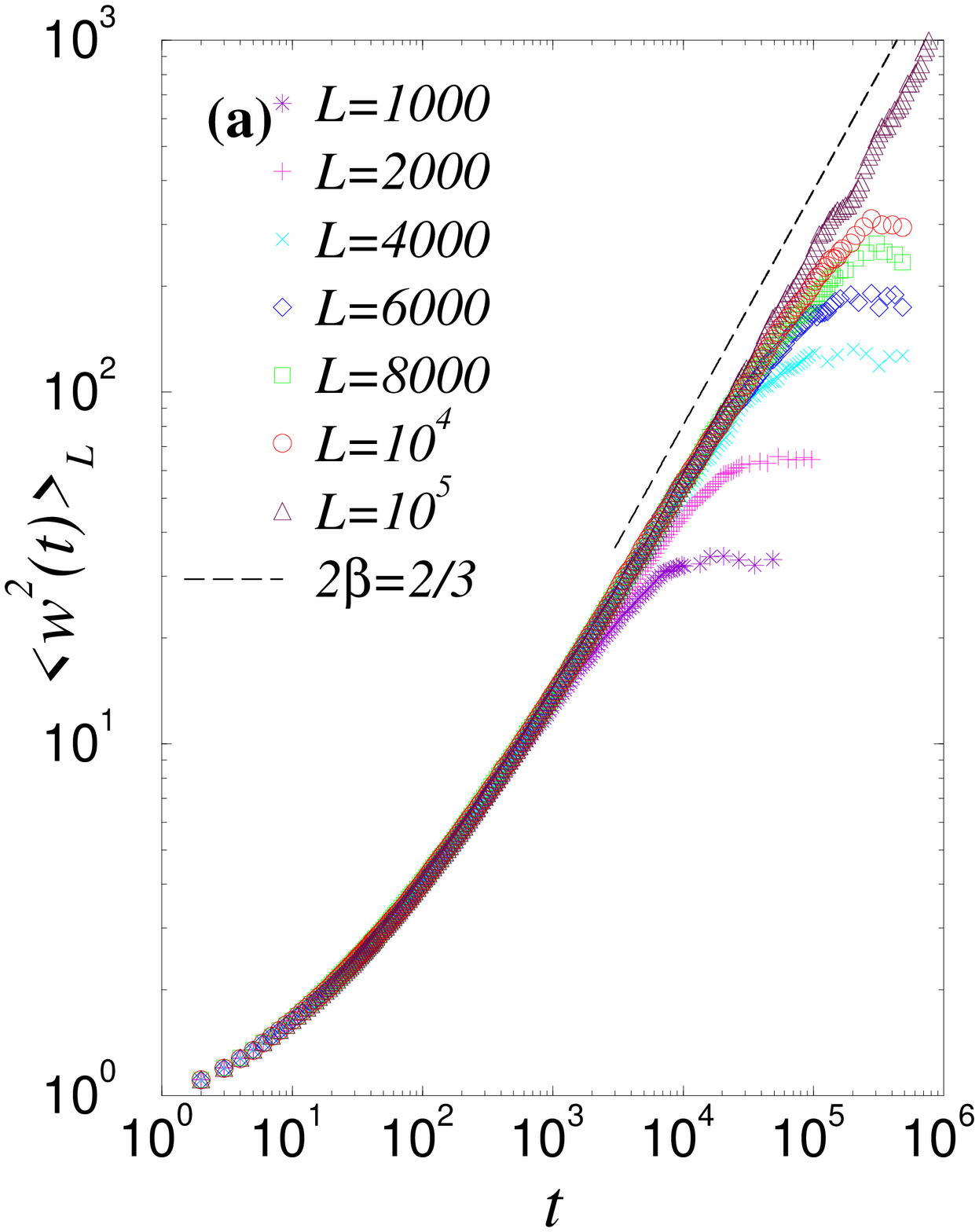}
\includegraphics[width=.35\textwidth]{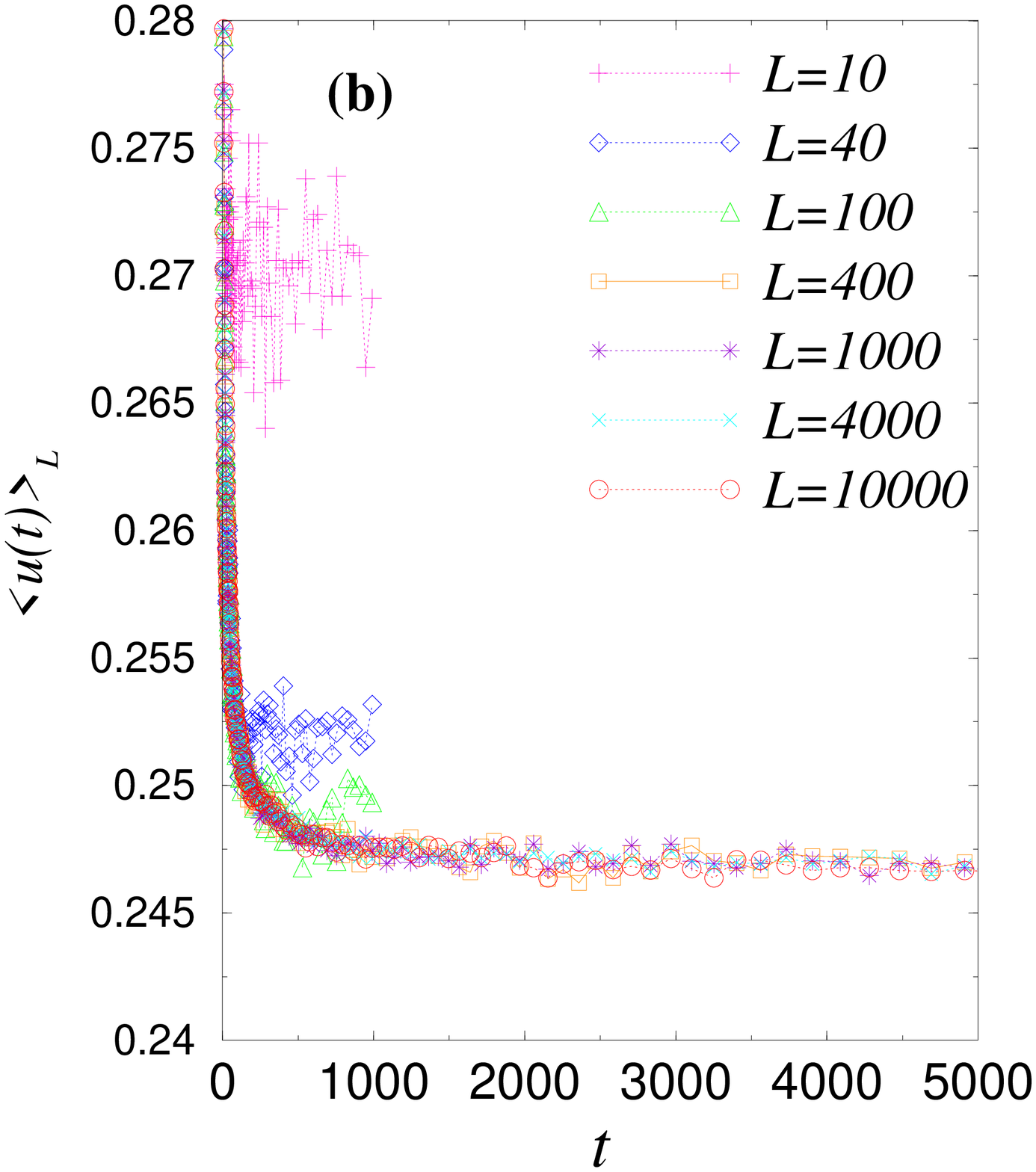}
\includegraphics[width=.32\textwidth]{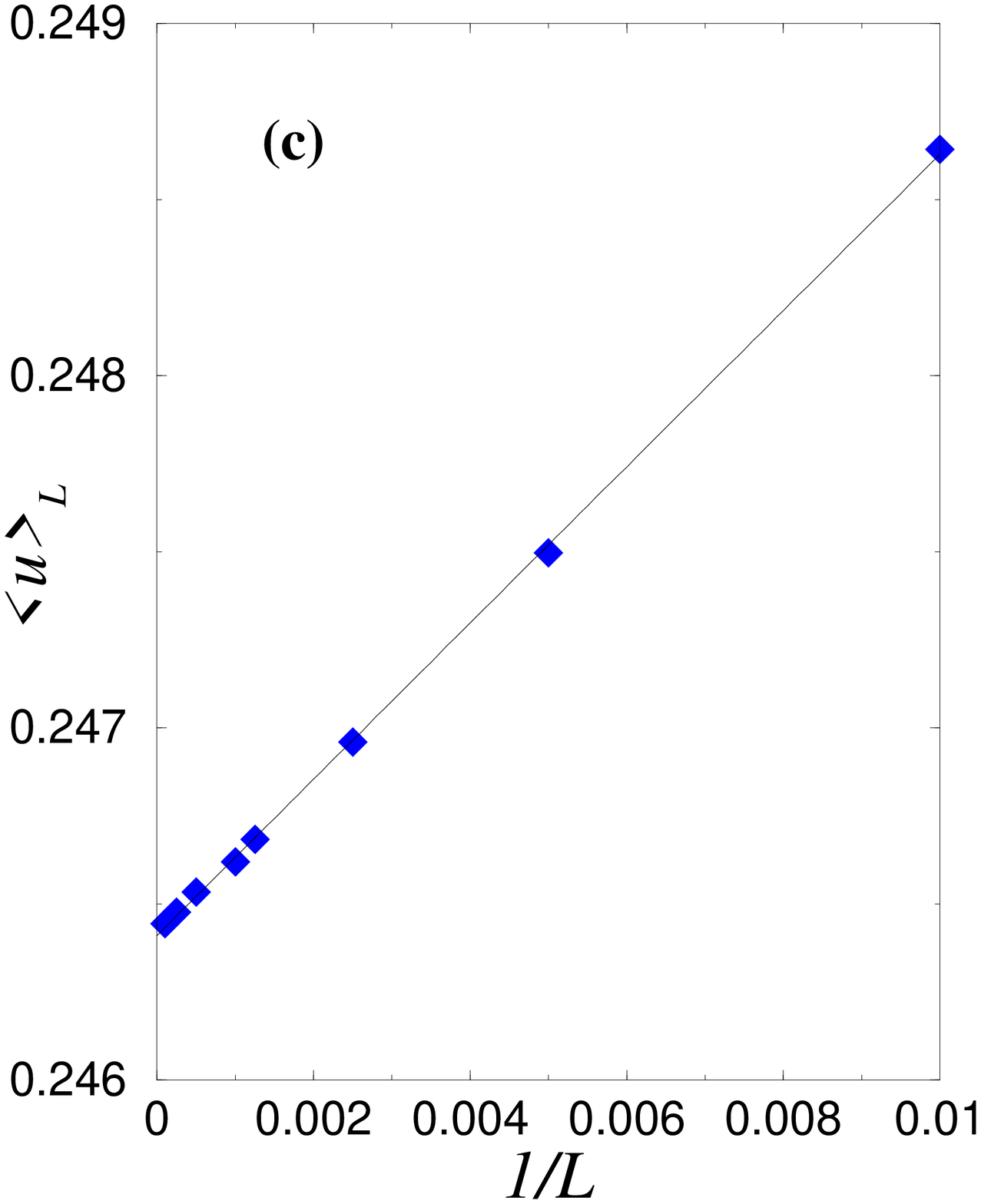}
\caption[]{Evolution of the simulated time horizon in one dimension:
(a) surface width; (b) density of local minima; (c) Steady-state scaling
of the parallel efficiency (density of local minima). Note the small vertical 
scales in (b) and (c)}
\label{fig1}
\end{figure}

For $d$$=$$1$ we showed \cite{KTNR} by coarse-graining and 
direct simulation of (\ref{tau_evolution}) that the evolution of the simulated
time horizon belongs to the KPZ universality class \cite{KPZ}. Our simulation
confirmed that before reaching the steady state, 
$\langle w^2(t)\rangle$$\sim$$t^{2\beta}$  with $\beta$$\approx$$1/3$  
[Fig.\ 1(a)]. At the same time the density of local minima, 
$\langle u(t)\rangle_L$, decreases monotonically with time towards
a long-time asymptotic limit well separated from zero [Fig.\ 1(b)].
The steady state is governed by the Edwards-Wilkinson 
Hamiltonian \cite{EW}, and the stationary width scales as 
$\langle w^2\rangle$$\sim$$L^{2\alpha}$, where $\alpha$$=$$1/2$ is the
roughness exponent. This guarantees that the coarse-grained landscape
is a simple random-walk surface; the local {\em slopes} are {\em short-range} 
correlated. Thus, the density of local minima is {\em non-zero}.
The non-zero density of local minima is a universal characteristic of this
class \cite{KTNR,TKSZ}. Further, its {\em steady-state} finite-size effects 
can be written as 
$\langle u \rangle_L = \langle u \rangle_{\infty} + {\rm const}/L$.
The ${\cal O}(1/L)$ correction in $d$$=$$1$ appears to 
be rather robust for periodic boundary conditions \cite{TKSZ}.
The extrapolated value for the efficiency is 
$\langle u \rangle_{\infty}$$=$$0.2464(1)$ [Fig.\ 1(c)].

\begin{figure}[t]
\includegraphics[width=.45\textwidth]{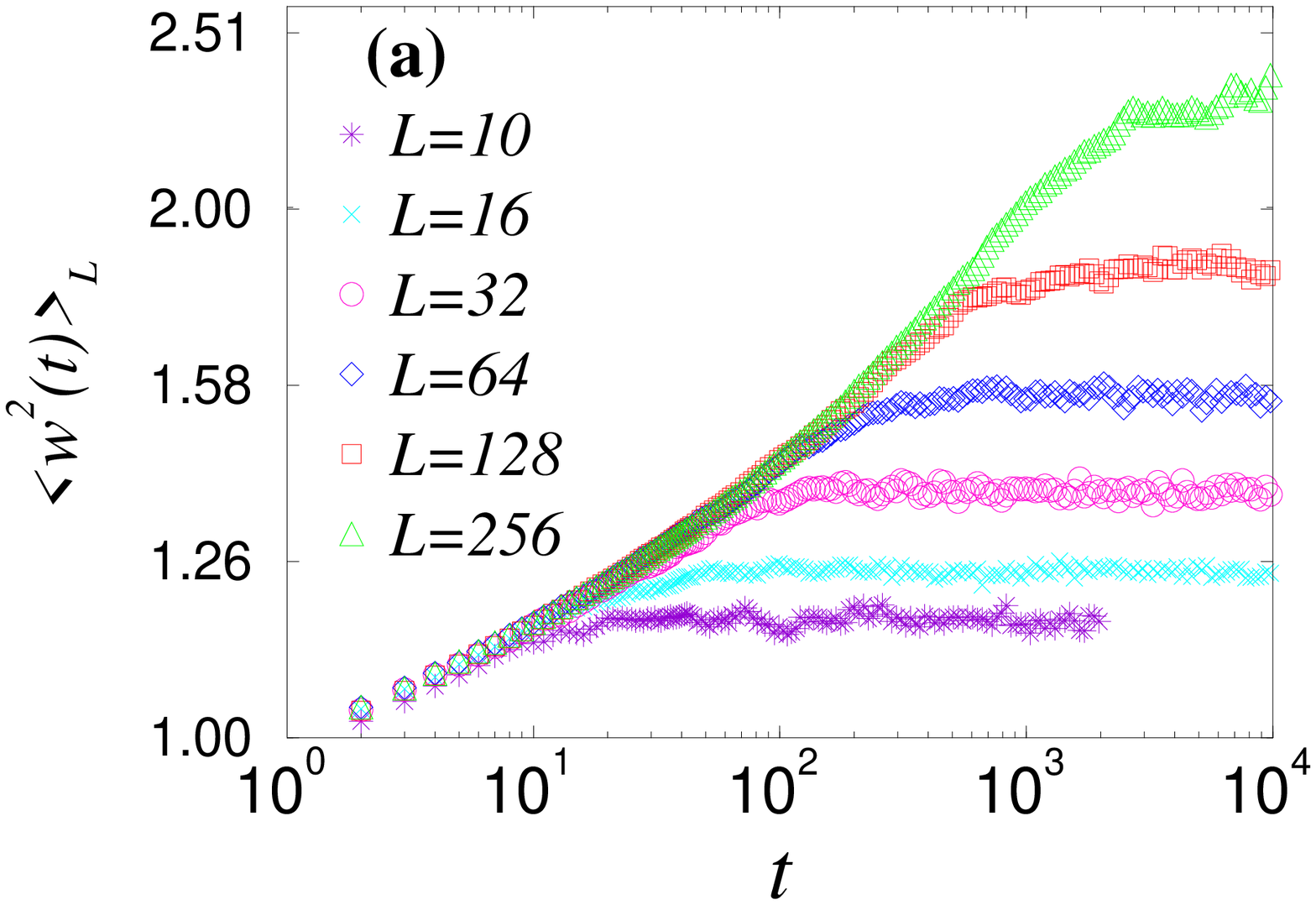}\hspace*{0.2truecm} 
\includegraphics[width=.45\textwidth]{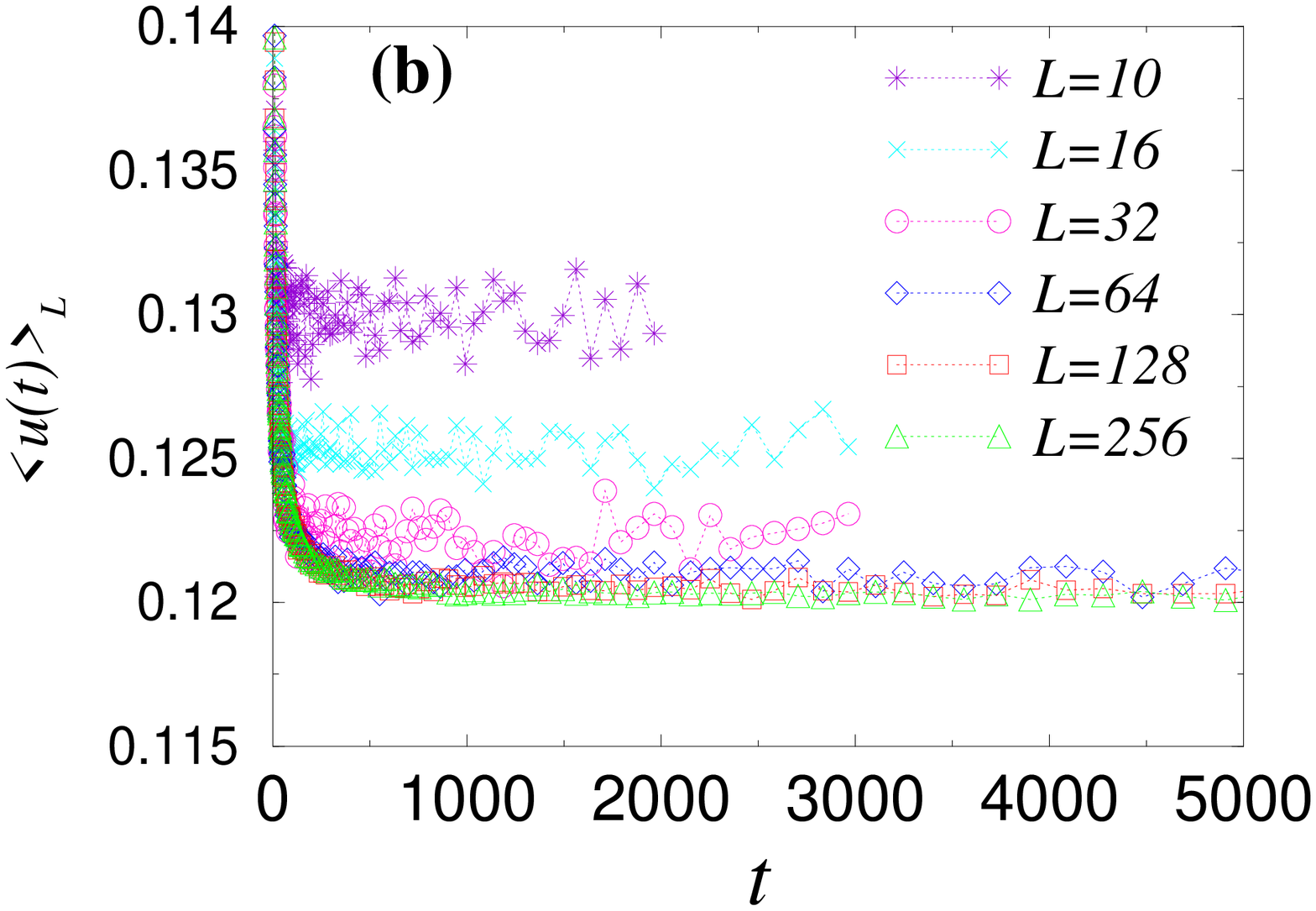} \\
\includegraphics[width=.45\textwidth]{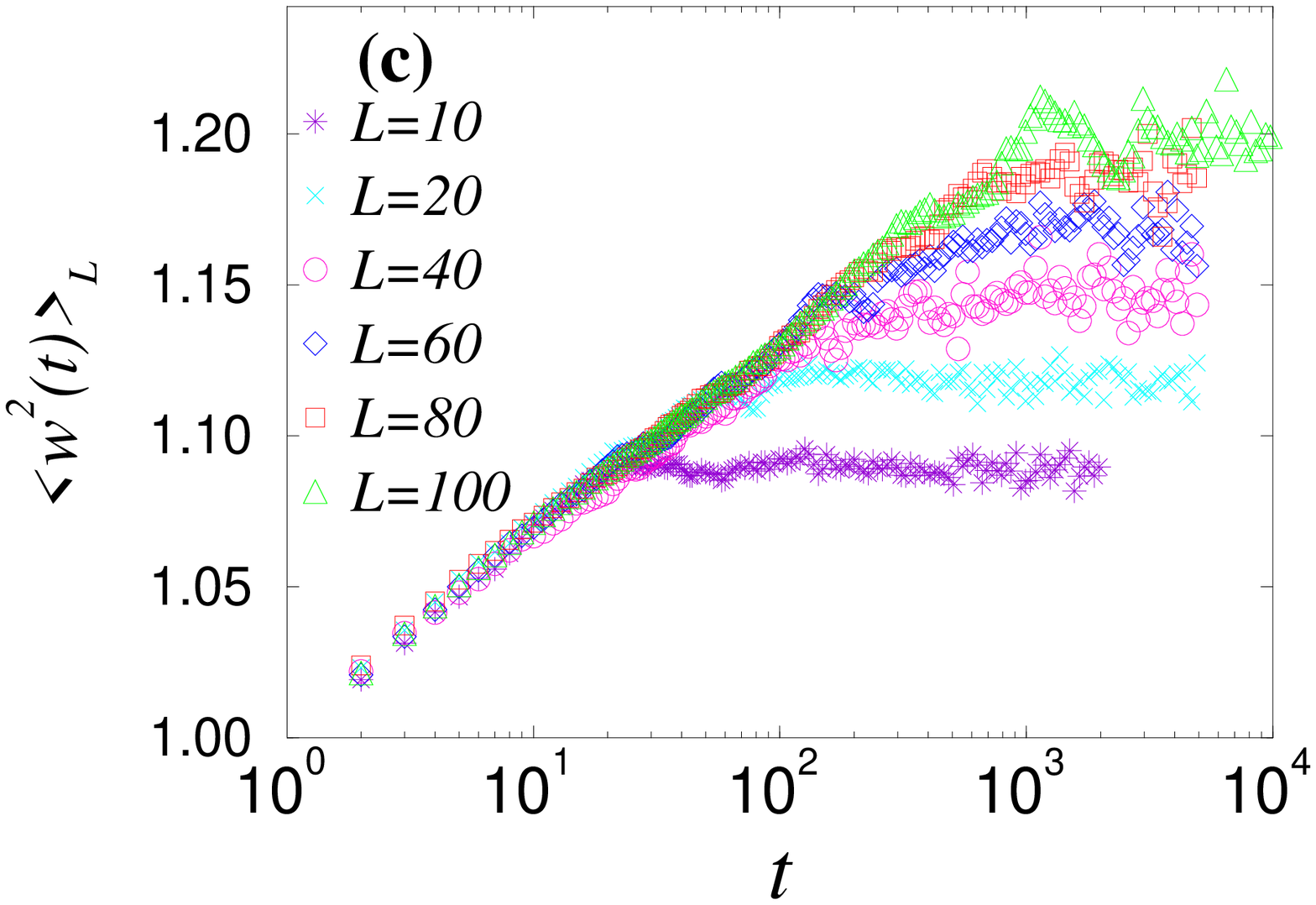}\hspace*{0.2truecm}
\includegraphics[width=.45\textwidth]{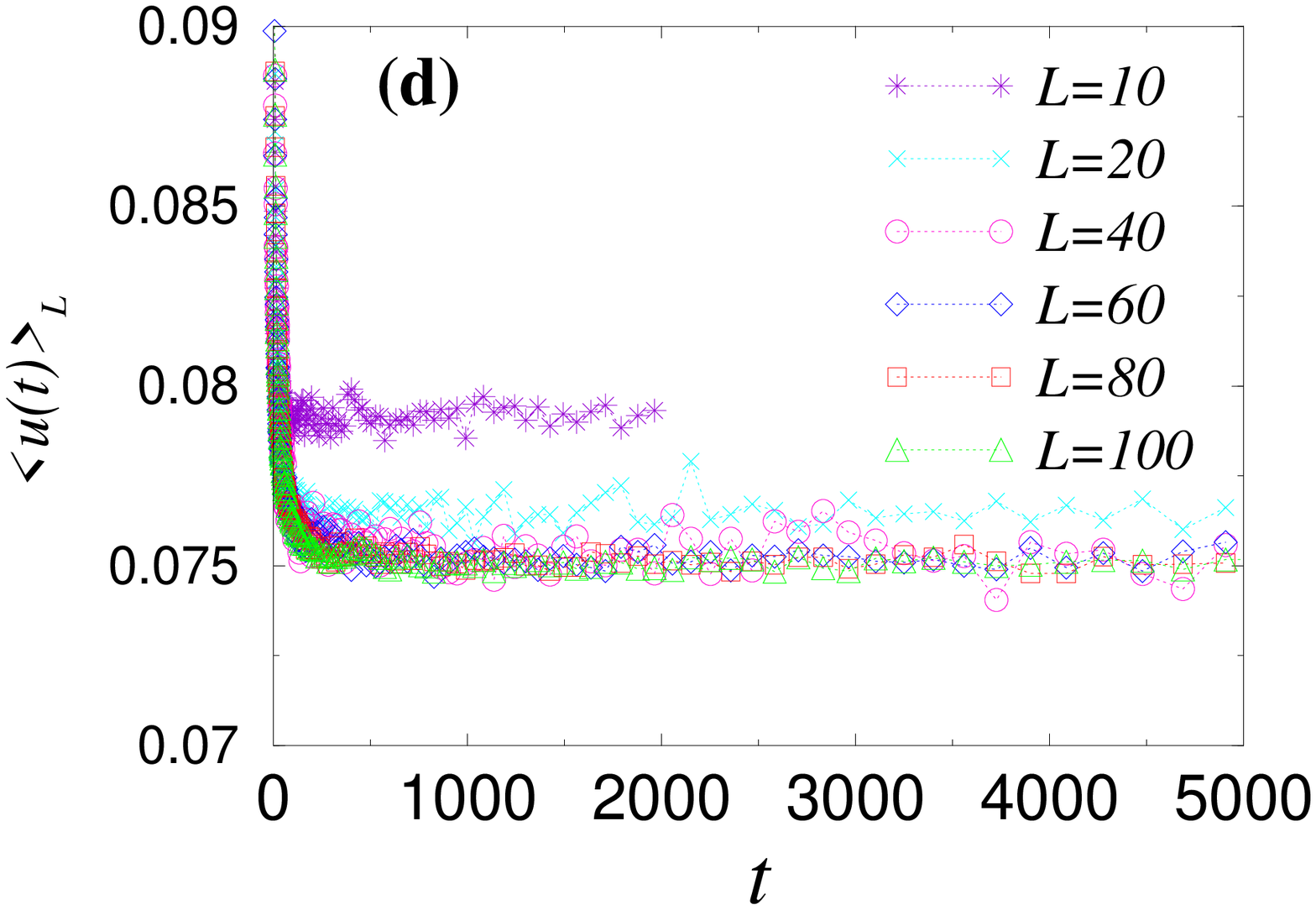}
\caption[]{Evolution of the simulated time horizon in $d$$=$$2$ and $d$$=$$3$:
(a) surface width in $d$$=$$2$; (b) density of local minima in $d$$=$$2$;
(c) surface width in $d$$=$$3$; (d) density of local minima in $d$$=$$3$.}
\label{fig3}
\end{figure}

In higher dimensions we observe the same qualitative behavior as for
$d$$=$$1$. The surface roughens and {\em saturates} for any finite
system, as seen in Fig.\ 2(a,c). 
Simultaneously, the density of local minima  decreases monotonically
towards its asymptotic ($t$$\rightarrow$$\infty$) finite-size value 
[Fig.\ 2(b,d)]. Again,
the steady-state density of local minima appears to be well 
separated from zero. For $d$$=$$2$ 
$\langle u \rangle_{\infty}$$\approx$$0.12$, and for $d$$=$$3$ 
$\langle u \rangle_{\infty}$$\approx$$0.075$. The  
$\langle u \rangle_{\infty}$$\sim$${\cal O}(1/K)$ behavior appears to be rather
general \cite{GSS}, where $K$$=$$2d$ is the number of nearest neighbors on a 
regular lattice.

Similar to the $d$$=$$1$ case, corrections to scaling are very strong,
both for the surface width and the density of local minima. While for $d$$=$$1$
we were able to simulate large systems ($L$$\gg$$10^3$) to obtain the KPZ
scaling exponents and the steady-state finite-size behavior of 
$\langle u \rangle_L$, in higher dimensions the relatively small system sizes
prevented us from extracting the scaling behavior of the width and the 
finite-size effects of the density of local minima.

We conjecture that the simulated time horizon exhibits
KPZ-like evolution in higher dimensions as well. While the 
scalability of the parallel scheme for random topology has been investigated 
earlier \cite{GSS}, the underlying mechanism for regular lattices (macroscopic
roughening) was only recently pointed out \cite{KTNR}. The major 
implication is that while the algorithm is scalable, and the width of the 
simulated time horizon saturates for any {\em finite} number of PEs,
it {\em diverges} in the limit of an infinite number of PEs. Thus, in an 
actual implementation the programmer has to take some actions to handle 
statistics collection.

\section{Summary and Outlook}
The analogy between the evolution of the simulated time horizon and 
non-equilibrium surface growth illustrates that when devising parallel 
algorithms, the programmer has to be very much concerned with the 
morphological properties of the associated surface. To fully describe the 
evolution of the simulated time horizon and the efficiency of the 
algorithm, one must focus on both the long wave-length behavior 
(the macroscopic width) and short-distance properties (the density of local 
minima). While most previous surface studies focus on the long 
wave-length properties, physics at short distances is just as challenging,
has real applications, and may reveal universal features 
\cite{KTNR,TKSZ}.

\section*{Acknowledgments}
We thank B.~D. Lubachevsky, A. Weiss, and S. Das Sarma for useful 
discussions. We acknowledge the support of DOE through SCRI-FSU, NSF-MRSEC at 
UMD, and NSF through Grant No. DMR-9871455.

%

\end{document}